\documentclass[a4paper]{jpconf}
\usepackage{amsfonts,amsmath,amssymb,cite,graphicx,slashed}
\usepackage[utf8]{inputenc}

\def\micromegas{{\sc MicrOMEGAs}}
\def\pythia{{\sc Pythia}}
\def\delphes{{\sc Delphes}}

\def\madanalysis{{\sc MadAnalysis}}
\def\feynrules{{\sc FeynRules}}

\def\ie{{\it i.e.}}
\newcommand{\be}{\begin{equation}}
\newcommand{\ee}{\end{equation}}
\def\bsp#1\esp{\begin{split}#1\end{split}}

\begin{document}
\title{Cornering top-philic dark matter with colliders and cosmology: the
  importance of QCD corrections}

\author{Benjamin Fuks$^{1,2}$}
\address{
  $^1$ Laboratoire de Physique Th\'eorique et Hautes Energies (LPTHE),
  UMR 7589, Sorbonne Universit\'e et CNRS, 4 place Jussieu,
  75252 Paris Cedex 05, France}
\address{
  $^2$ Institut Universitaire de France, 103 boulevard Saint-Michel, 75005
  Paris, France}
\ead{fuks@lpthe.jussieu.fr}

\begin{abstract}
Constraining dark matter models necessitates accurate predictions for large set
of observables originating from collider physics, cosmology and astrophysics. We
consider two classes of top-philic dark matter models where the dark sector is
coupled to the Standard Model via the top quark and study the complementarity of
dark matter relic density, direct and indirect detection and collider searches
in exploring the model parameter space. We moreover investigate how
higher-order corrections affect the results.
\end{abstract}

\section{Introduction}
Dark matter is convincingly evidenced by numerous measurements and observations,
like the cosmic microwave background, the flattening of the galaxy rotation
curves or gravitational lensing. These many hints triggered a huge endeavour to
detect dark matter and study its properties in experiments that range from
direct and indirect probes to collider searches. However, despite this huge
experimental effort, there is currently no indication about the true nature of
dark matter. As a consequence, dark matter scenarios are more and more
constrained by many complementary datasets. We illustrate the impact of this
complementarity for models in which dark matter is top-philic, \ie~when it
dominantly couples to top quarks~\cite{Arina:2016cqj,Colucci:2018vxz}. We
moreover assess the impact of QCD higher-order corrections on the predictions,
both for collider processes and dark matter annihilation in the universe. In
Section~\ref{sec:cosmo}, we focus on cosmological constraints, whilst
Section~\ref{sec:colliders} is dedicated to collider constraints. We conclude by
combining these two sets of constraints in Section~\ref{sec:combine}.

\section{Probing top-philic dark matter with cosmology}\label{sec:cosmo}
For concreteness, we consider a simplified top-philic dark matter scenario in
which the dark matter particle is a scalar $S$ of mass $m_S$ and interacts with
the top quark $t$ through a Yukawa interaction $\tilde y_t$ involving a
vector-like quark $T$ of mass $m_T$. We moreover impose a $\mathbb{Z}_2$
discrete symmetry under which all Standard Model fields are even and new physics
states odd, which guarantees dark matter stability and forbids any mixing of
the heavy quark with the top quark. This model is described by the Lagrangian
\be
 {\cal L} = {\cal L}_{\rm SM}
  + i \bar T \slashed{D} T - m_T \bar T T
  +  \frac 12 \partial_\mu S \partial^\mu S - \frac 12 m_S S^2
  + \Big[ \tilde{y}_t\, S\ \bar T P_R t + {\rm h.c.} \Big] \ ,
\ee
where $P_R$ denotes the right-handed chirality projector, ${\cal L}_{\rm SM}$
the Standard Model Lagrangian, and where any coupling of dark matter to the
Higgs sector is considered vanishing. The model parameter
space is then defined by three parameters that we choose to be the dark matter
mass $m_S$, the compression factor of the spectrum $r-1 = m_T/m_S-1$ and the
Yukawa coupling $\tilde{y}_t$.

\begin{figure}
  \begin{center}
     \includegraphics[width=.48\textwidth]{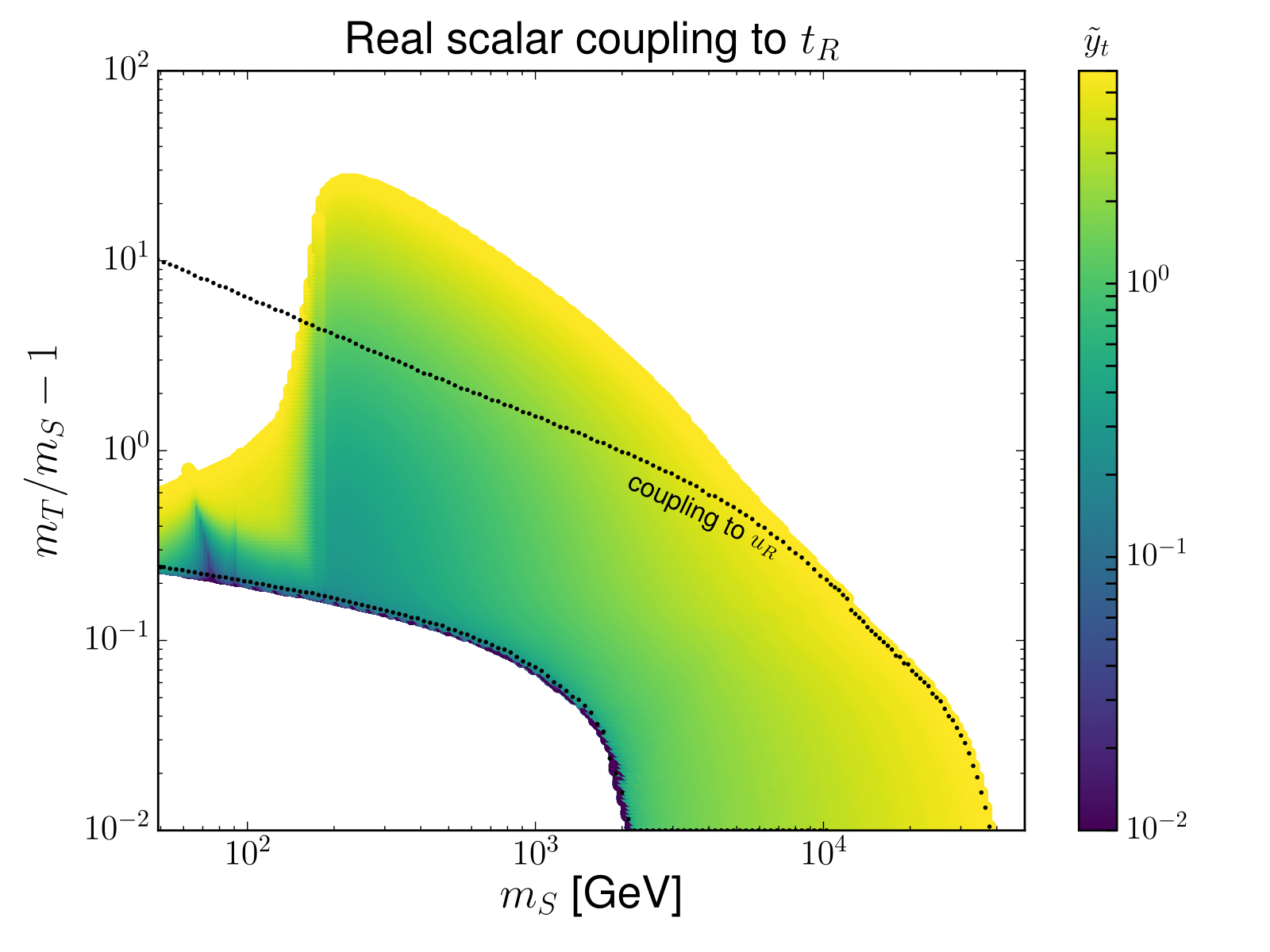}
     \includegraphics[width=.48\textwidth]{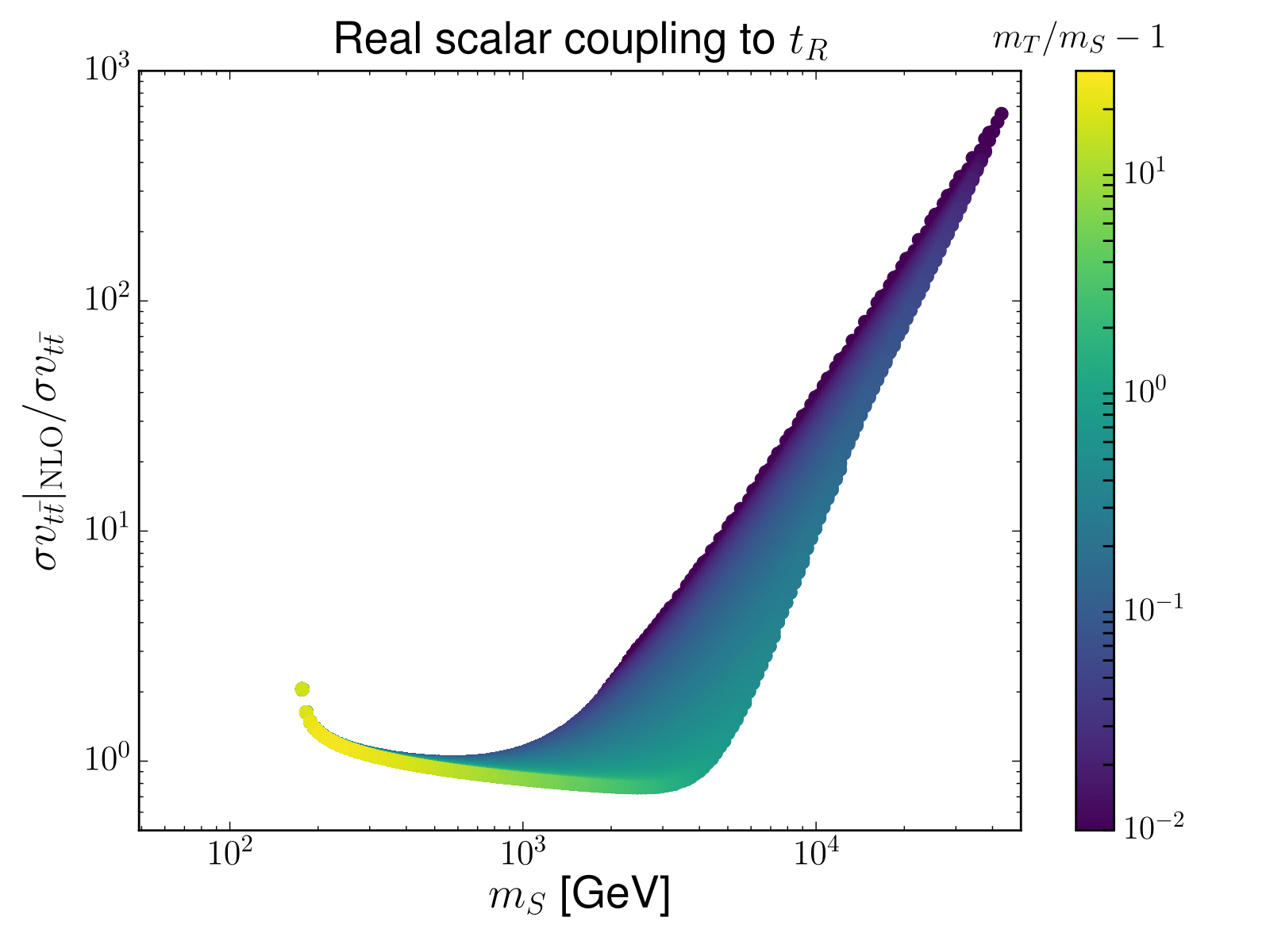}
    \caption{Left: Region of the parameter space in which the measured relic
      abundance can be accommodated by fixing $\tilde y_t$ (given through the
      colour code). The results are presented in the $(m_S,r-1)$ plane. For
      comparison, the dotted black contour represents the viable parameter space
      region obtained when the top mass is neglected. Right: Ratio of the NLO
      annihilation cross section into top-antitop pairs to the LO one.}
  \label{fig:relic}
  \end{center}
\end{figure}

\subsection{Relic abundance}\label{sec:relic}
In order to determine the dark matter relic abundance, we assume standard
cosmology and a thermal freeze-out mechanism to describe dark matter dynamics in
the early universe. The annihilation cross section is evaluated by
including all tree-level $2\to2$ diagrams, to which we supplement contributions
from dark matter annihilations into a three-body $tWb$ system, Sommerfeld
enhancement, annihilations into gluons and photons as well as
next-to-leading-order (NLO) QCD corrections to the annihilation cross section
into top-antitop pairs. Whilst threshold corrections should in principle be
included as affecting the production of a top-antitop system at
rest~\cite{Drees:1990dq}, they are expected to yield subleading corrections due
to the absence of any toponium bound state. They are thus omitted. In the left
panel of Figure~\ref{fig:relic}, we present numerical predictions obtained with
\micromegas~\cite{Belanger:2014vza}, after having modified the program to
accomodate the above-mentioned features. We show the region of the $(m_S, r-1)$
plane for which there exists a value of the Yukawa coupling $\tilde y_t$,
indicated through the colour code, that yields a relic density $\Omega h^2=0.12$
compatible with Planck data~\cite{Ade:2013zuv}. Our results ignore
scenarios featuring a $\tilde y_t$ coupling smaller than $10^{-4}$ and larger
than 6, the former being associated with a too naive treatment of
co-annihilations by \micromegas~\cite{Garny:2017rxs} and the latter
corresponding to a potential breakdown of perturbation theory.

For very heavy dark matter with $m_S>5~{\rm TeV}$, dark matter
annihilations into top quarks dominate. The top mass $m_t$ however plays a
subleading role, the viable region of the parameter space (coloured area)
matching the contour obtained when $m_t$ is neglected (dotted black). In this
kinematical regime, the annihilation cross section is entirely driven by NLO
effects, as illustrated on the right panel of Figure~\ref{fig:relic} where
we present $K$-factors defined as the ratio of the NLO result to the
leading-order (LO) one for all viable scenarios. The huge $K$-factors
representative of the $m_S>5~{\rm TeV}$ region are driven by virtual internal
bremsstrahlung (VIB) diagrams~\cite{Giacchino:2013bta,Toma:2013bka,%
Colucci:2018qml}. Final-state
radiation is indeed helicity suppressed (as proportional to $m_t/m_S$), in
contrast to VIB emission that is proportional to $m_T/m_S$ and thus enhanced as
the spectrum is compressed in this parameter space region (see the colour code
in the figure).
For moderate dark matter masses $m_t <m_S <5$~TeV, annihilations into top quarks
still dominate but the behaviour is this time driven by tree-level $S$-wave
contributions, the NLO corrections being small ($K$-factors of ${\cal O}(1)$).
In the massless limit, annihilations into quark pairs are negligible and the
relic density is driven by loop-induced annihilations into gluons (dotted black
contour in the left panel of Figure~\ref{fig:relic}). In contrast, in the
massive case, significant annihilations into quarks allow for larger $r-1$
choices to accommodate the measured relic density (coloured contour).
In the light dark matter case ($m_S < m_t$), the relic density is governed by
annihilations into a $tWb$ system or gluons, or through co-annihilations with
the mediator for more compressed scenarios. Co-annihilations play an important
role near $m_T+m_S\simeq m_t$ as they are resonantly enhanced, which corresponds
to the dark blue region in the left panel of Figure~\ref{fig:relic} ($m_S\sim
70-80$~GeV). Annihilations into gluons are only relevant when all
co-annihilation channels are closed (\ie~for very light dark matter), and
annihilations into a $tWb$ system only contribute close to threshold for $m_S\in
[(m_t+m_W)/2, m_t]$.

\subsection{Direct detection}
\begin{figure}
  \begin{center}
     \includegraphics[width=.48\textwidth]{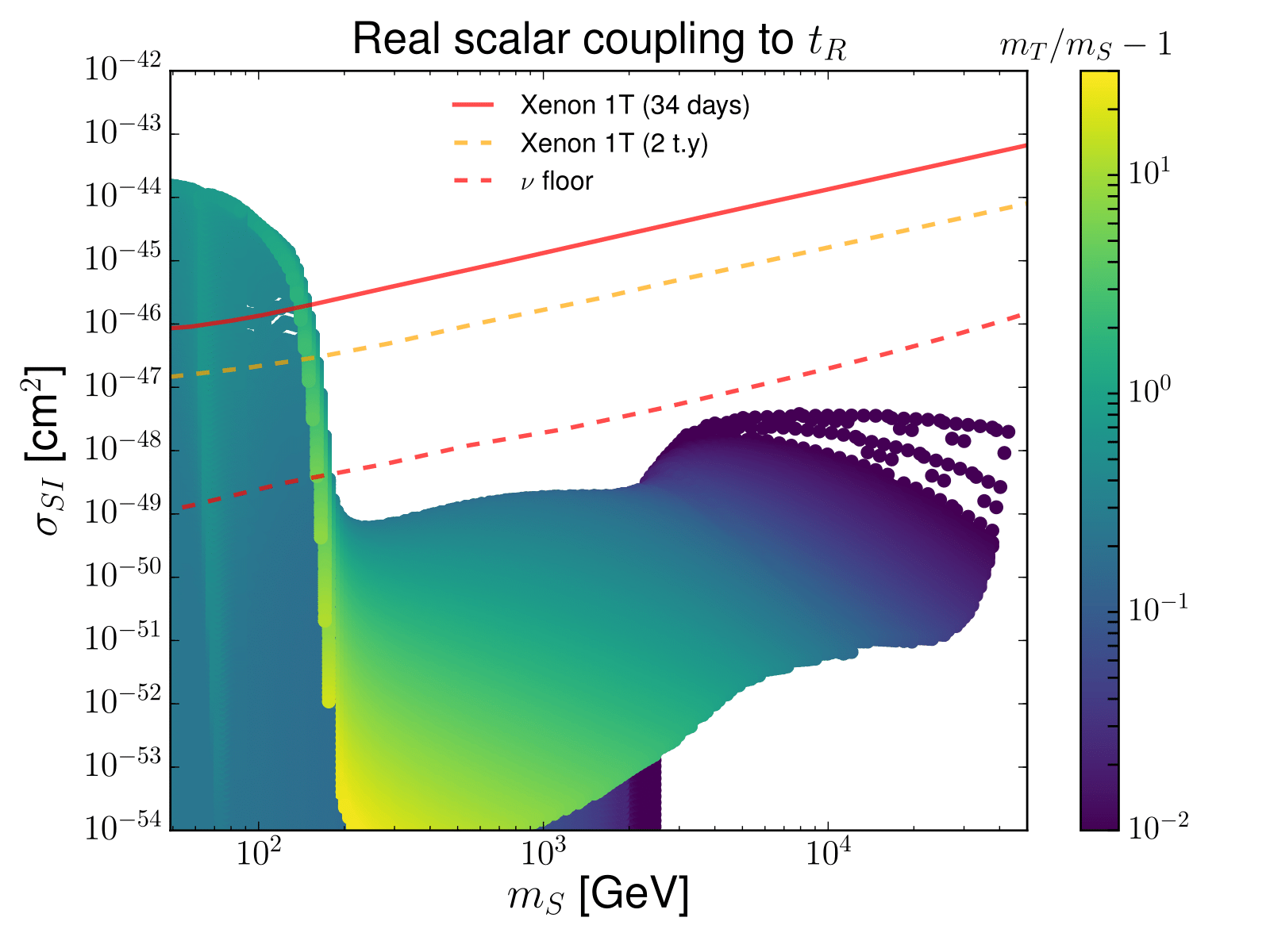}
    \caption{Spin-independent dark matter scattering cross section as a function
      of the dark matter mass $m_S$. For each scenario, the $\tilde y_t$ value
      is set to reproduce the observed relic density and the compressibility
      factor is depicted by the colour code. We indicate current (solid red) and
      future (dashed orange) 90\% confidence level exclusions by the Xenon 1T
      experiment~\cite{Aprile:2017iyp,Aprile:2015uzo}. The dashed red line
      corresponds to the neutrino floor.}
  \label{fig:DD}
  \end{center}
\end{figure}

In the scenario studied in this section, dark matter scattering off nuclei
proceeds via loop-induced interactions with the gluon field strength tensor
$G_{\mu\nu}$, that could be modelled
effectively~\cite{Hisano:2015bma},
\be
 \mathcal{L}_{g} =  \frac{\tilde y_t^2}{48 \big[m_T^2-m_S^2\big]}
    \frac{\alpha_s}{\pi} S^2 G^{\mu \nu}_a G_{\mu \nu}^a \ .
\ee
Accounting for a nucleus of mass $m_A$ and comprised of $Z$ protons $p$ and
$(A-Z)$ neutrons $n$, the spin-independent dark matter scattering cross section
$\sigma_{\rm SI}$ is derived from the coherent sum of the proton and the neutron
contributions,
\be
  \sigma_{\rm SI}=\frac{m_A^2}{\pi(m_S+m_A)^2} \bigg[Z f_p + (A-Z)f_n\bigg]^2
  \quad\text{with}\quad
  \frac{f_N}{m_N}= -\frac{\tilde y_t^2}{54 \big[m_T^2-m_S^2\big]}
      \bigg[1-\sum_{q=u,d,s}f_{T_q}^{(N)}\bigg] \ .
\ee
This expression depends on the coupling $f_N$ of $S$ to a nucleon $N$ of mass
$m_N$, that is related to the quark mass fractions $f_{T_q}^{(N)}$ deduced from
lattice calculations~\cite{Oksuzian:2012rzb}.

In Figure~\ref{fig:DD}, we present the $m_S$-dependence of $\sigma_{\rm SI}$ for
all viable scenarios featuring a relic density in agreement with Planck data.
Below the top threshold, the largest cross section values are obtained for
setups in which the relic density is driven by annihilations into gluons and for
which $\tilde y_t$ is large. Above the top threshold, the $\tilde y_t$ value
necessary to get the right relic abundance drops, and so does $\sigma_{\rm SI}$.
On different lines, $St \to Tg$ co-annihilations impact the results for $m_S
\gtrsim2.5$~TeV, yielding an enhancement of the scalar interactions with gluons
in ${\cal L}_g$. For most scenarios, $\sigma_{\rm SI}$ lies below the neutrino
floor and is thus not reachable by direct detection experiments. For spectrum
featuring light dark matter annihilating into gluons, many scenarios are however
within the present (34 days of exposure) and future (2.1 years of data
acquisition) reach of Xenon 1T~\cite{Aprile:2017iyp,Aprile:2015uzo}.

\subsection{Indirect detection}

\begin{figure}
  \begin{center}
    \includegraphics[width=0.48\textwidth]{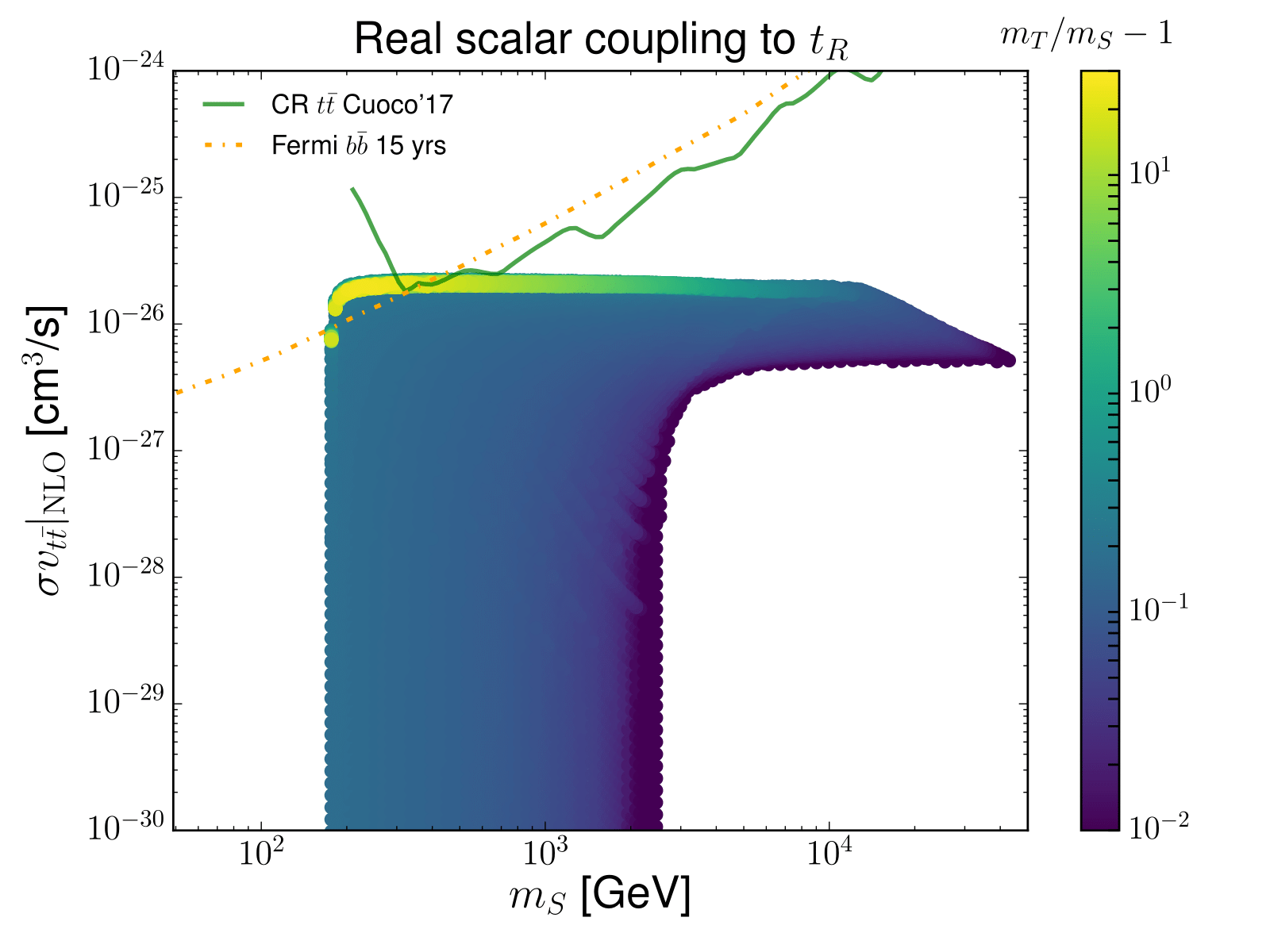}
    \includegraphics[width=0.48\textwidth]{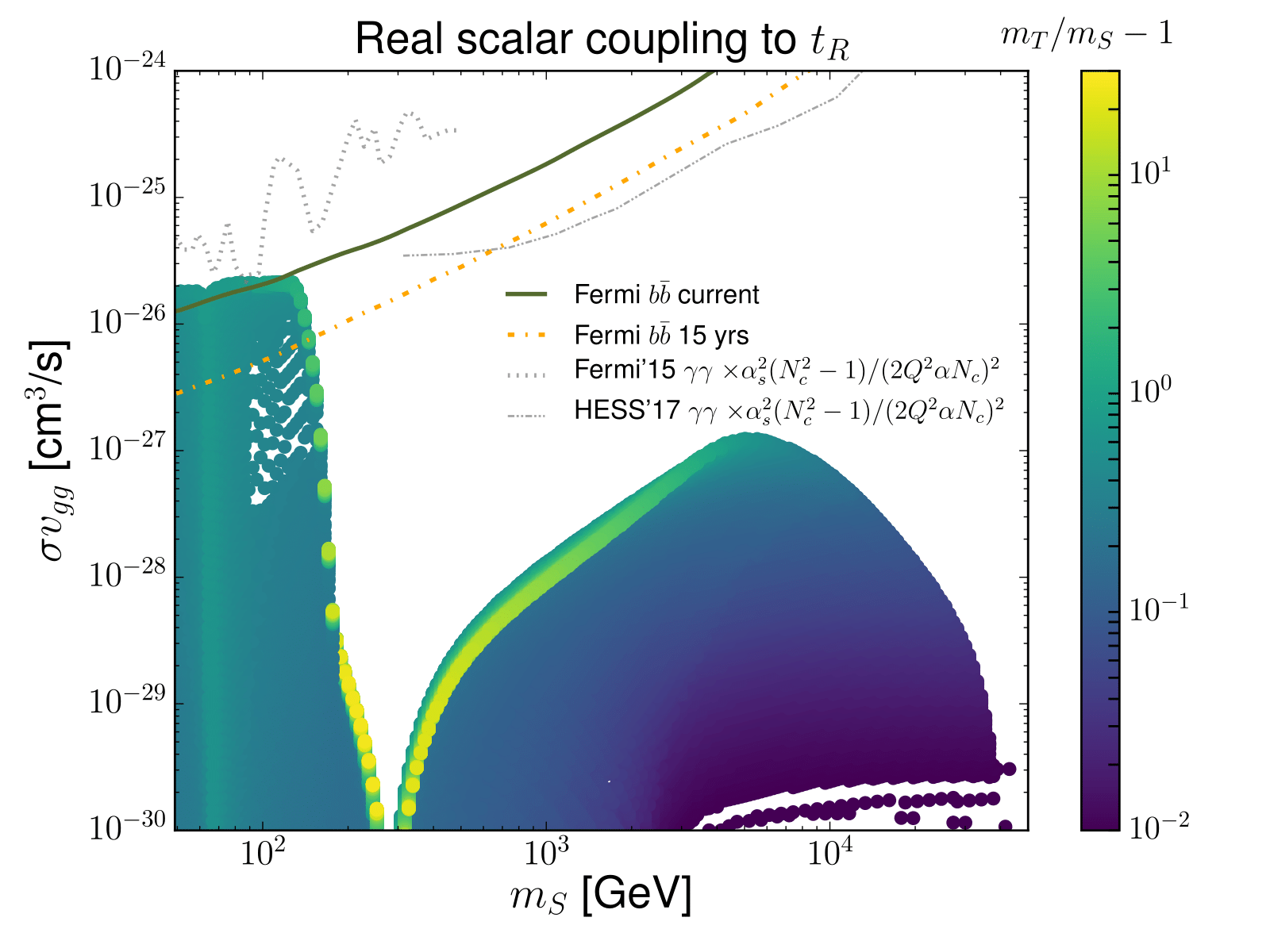}
    \caption{NLO $SS \to t\bar t$ (left) and loop-induced LO $SS\to gg$ (right)
      annihilation cross sections at zero velocity. We indicate the limits
      obtained from cosmic rays (solid light green), and from the analysis of
      current (solid dark green) and future (dot-dashed orange) Fermi-LAT dwarf
      spheroidal galaxy data in the $b\bar b$ channel. Upper limits obtained
      from the gamma-ray analysis of H.E.S.S.~(double-dotted dashed grey) and
      Fermi-LAT (dotted grey) are also shown. For each scenario, the $\tilde
      y_t$ value is set to reproduce the observed relic density and the
      compressibility factor is depicted by the colour code.}
    \label{fig:ID}
  \end{center}
\end{figure}

In Figure~\ref{fig:ID} we present, for all viable scenarios, predictions for the
dark matter annihilation cross section at zero velocity for a top-antitop
($\sigma v_{t\bar t}$, left panel) and gluon-pair ($\sigma v_{gg}$, right panel)
final state. The respective cross sections are evaluated at the NLO and
loop-induced LO accuracy in QCD. As already found in the previous subsections,
annihilations into gluons are only relevant below the top threshold, and
annihilations into top-antitop pairs dominate for $m_S > m_t$.
We moreover recall that NLO effects largely affect the predictions for
$m_S>5$~TeV (see Section~\ref{sec:relic}).

Dark matter annihilations into top-antitop pairs can be constrained from
antiproton cosmic ray fluxes~\cite{Cuoco:2017iax} (solid light green
line), and annihilations into both top quarks and gluons can be constrained by
the analysis of gamma-ray Fermi-LAT dwarf spheroidal data in the $b\bar b$
mode~\cite{Ackermann:2015zua} (solid dark green line). Similarly, the future
gamma-ray sensitivity to the model can be extrapolated after accounting for
the prospects of 15 years of Fermi-LAT running~\cite{Charles:2016pgz}
(dot-dashed orange lines). Although there is no direct Fermi-LAT results for
annihilations into the considered final states, the subsequent secondary
gamma-ray spectrum is expected to exhibit a similar behaviour as for
annihilations into a $b\bar b$ system~\cite{Colucci:2018vxz}. An estimate of the
limits can therefore be extracted through a reweighting
procedure~\cite{Bringmann:2012vr},
\be
  \sigma v_{gg, t\bar t} = \sigma v_{b \bar b}
    \frac{N_\gamma^{b\bar{b}}}{N_\gamma^{gg,t \bar t}} \ ,
\ee
where $N_\gamma^{X}$ is the number of photons originating from an $X$ final
state that we estimate from the hadronisation model implemented in the
\pythia~8 package~\cite{Sjostrand:2014zea}.

H.E.S.S.~\cite{Rinchiuso:2017kfn} (double-dot-dashed grey line) and Fermi-LAT
gamma-ray data~\cite{Ackermann:2015lka} (dashed grey line) can potentially
constrain the model when considering direct dark matter annihilations into
photons. The latter is expected to be enhanced in two regimes, \ie~for light
dark matter ($m_S<m_t$) where loop-induced annihilations into photons could be
important (most other channels being closed) and in the multi-TeV regime where
radiative emission is dominated by VIB diagrams. The corresponding cross
sections, associated with dark matter annihilations into photons and into a
$t\bar t\gamma$ system, can be deduced from
\be
  \sigma v_{\gamma\gamma} =
    \frac{4Q^4\alpha^2N_c^2}{\alpha_S^2\left(N_c^2-1\right)} \sigma v_{gg}
  \qquad\text{and}\qquad
  \sigma v_{t \bar t \gamma} =
    {2 N_c Q^2 \alpha \over (N_c^2-1)\alpha_s} \sigma v_{t \bar t g}\ ,
\ee
where $N_c=3$ denotes the number of colours, $\alpha$ and $\alpha_S$ the
electromagnetic and strong coupling, and $Q$ the electric charge of the heavy
quark $T$. The predictions can be confronted to the results of Fermi-LAT at
energies around and below $m_t$ as well as to the results of H.E.S.S. for larger
dark matter masses (following the procedure of Ref.~\cite{Ibarra:2013eda}).

As shown in Figure~\ref{fig:ID}, present and future gamma-ray and cosmic-ray
data turns out to be sensitive to only a small fraction of the model parameter
space.

\section{Collider constraints on top-philic dark matter}\label{sec:colliders}
In full generality, top-philic dark matter models can be probed at
colliders via several processes~\cite{Arina:2016cqj}. First, dark
matter can be produced from (virtual or resonant) mediator exchanges,
potentially together with one or more Standard Model particles. Whilst dark
matter is stable and leaves a missing energy signature in the detector, the
extra Standard Model objects are visible and provide handles on new physics.
Next, depending on the details of the model, mediator production can also
resonantly leads to a purely Standard Model final state, which offers hence
another set of probes of the model.

The dark matter model detailed in Section~\ref{sec:cosmo} does not give rise to
the most general phenomenology as it cannot be probed through resonance searches
in Standard Model final states. We therefore focus on a different model that can
be probed through both dark
matter searches and resonance searches. We illustrate this feature, as well as
the impact of the QCD corrections, by studying, for the sake of the example, LHC
Run~1 results. The final results on the model of the previous section will
nevertheless include LHC Run~2 data and will be shown in Section~\ref{%
sec:combine}. We consider a simplified top-philic dark matter model built by
supplementing the Standard Model with a Dirac fermionic dark matter field $\chi$
of mass $m_X$ and a scalar mediator $Y_0$ of mass $m_Y$. The interactions are
described by the Lagrangian
\be
  {\cal L} = {\cal L}_{\rm SM} + {\cal L}_{\rm kin}
     -\Big(g_t \frac{y_t}{\sqrt{2}} \bar{t}t + g_X \bar\chi \chi \Big) Y_0 \ ,
\ee
in which we denote by $g_t$ and $g_X$ the coupling strengths of the mediator to
the Standard Model and dark sector respectively. We have assumed that the
mediator couples to fermions with a strength proportional to the corresponding
Yukawa coupling $y_f$, so that only the top quark can be considered, and
${\cal L}_{\rm kin}$ includes new physics kinetic and mass terms.

For our collider simulations, we have implemented the above Lagrangian into
\feynrules~\cite{Alloul:2013bka}, which we have jointly used with NLOCT~\cite{%
Degrande:2014vpa} to generate an NLO UFO model~\cite{Degrande:2011ua}. We have
made use of MG5\_aMC@NLO~\cite{Alwall:2014hca} to generate hard-scattering
events obtained by the convolution of LO and NLO matrix elements with the
NNPDF~2.3 set of parton distribution functions~\cite{Ball:2012cx}. The resulting
events are matched with parton showers as modeled by \pythia~8~\cite{%
Sjostrand:2014zea}, that is also used to describe hadronisation. The response of
the LHC detectors is handled by \delphes~3~\cite{deFavereau:2013fsa} that is
tuned appropriately~\cite{Dumont:2014tja}, and signal efficiencies and exclusion
contours are estimated by reinterpreting the results of various LHC analyses
with \madanalysis~5~\cite{Conte:2012fm,Conte:2014zja,Conte:2018vmg}.

\subsection{Searches with missing transverse energy}
\begin{figure}
  \center
  \includegraphics[width=0.48\textwidth]{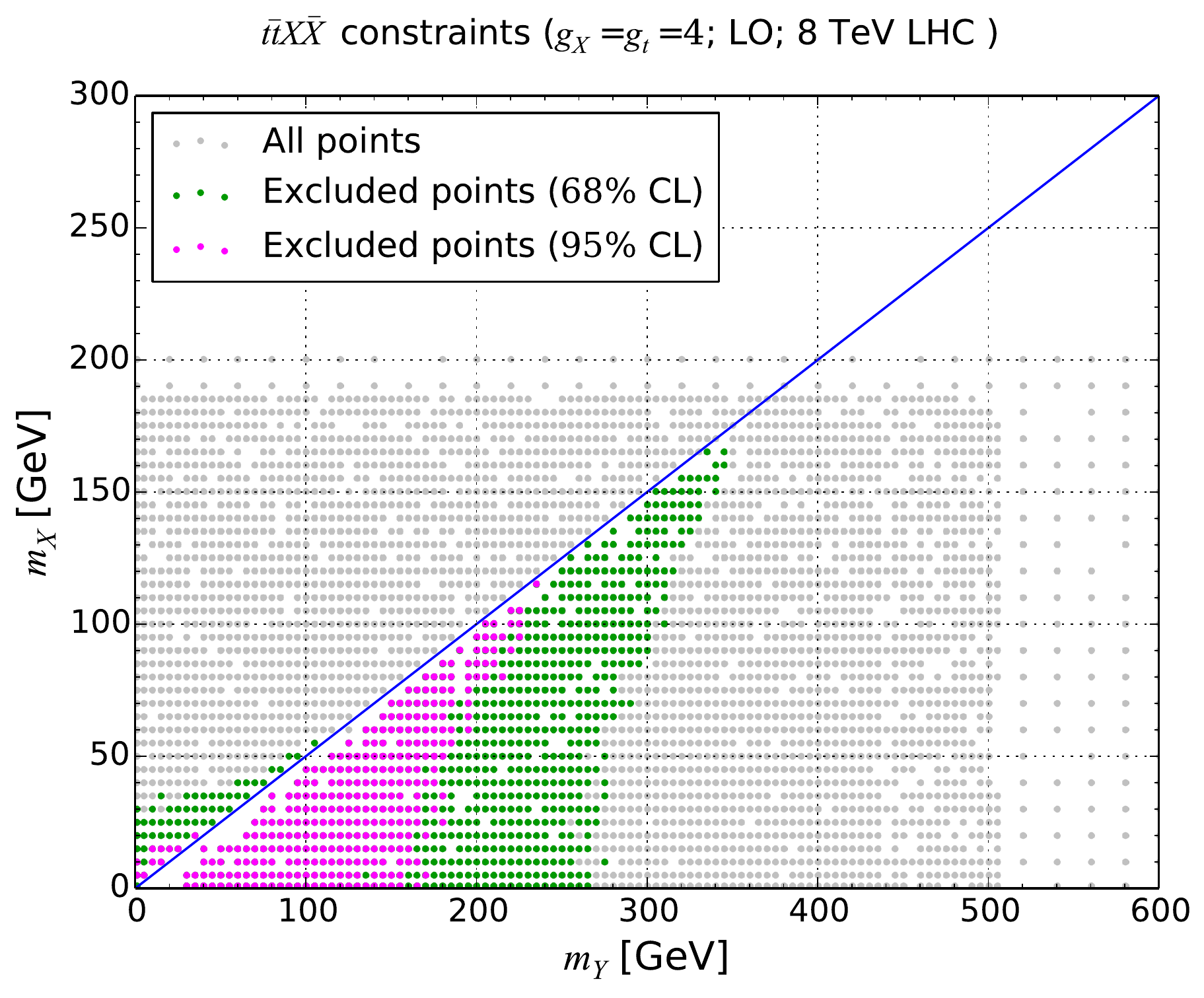}
  \includegraphics[width=0.48\textwidth]{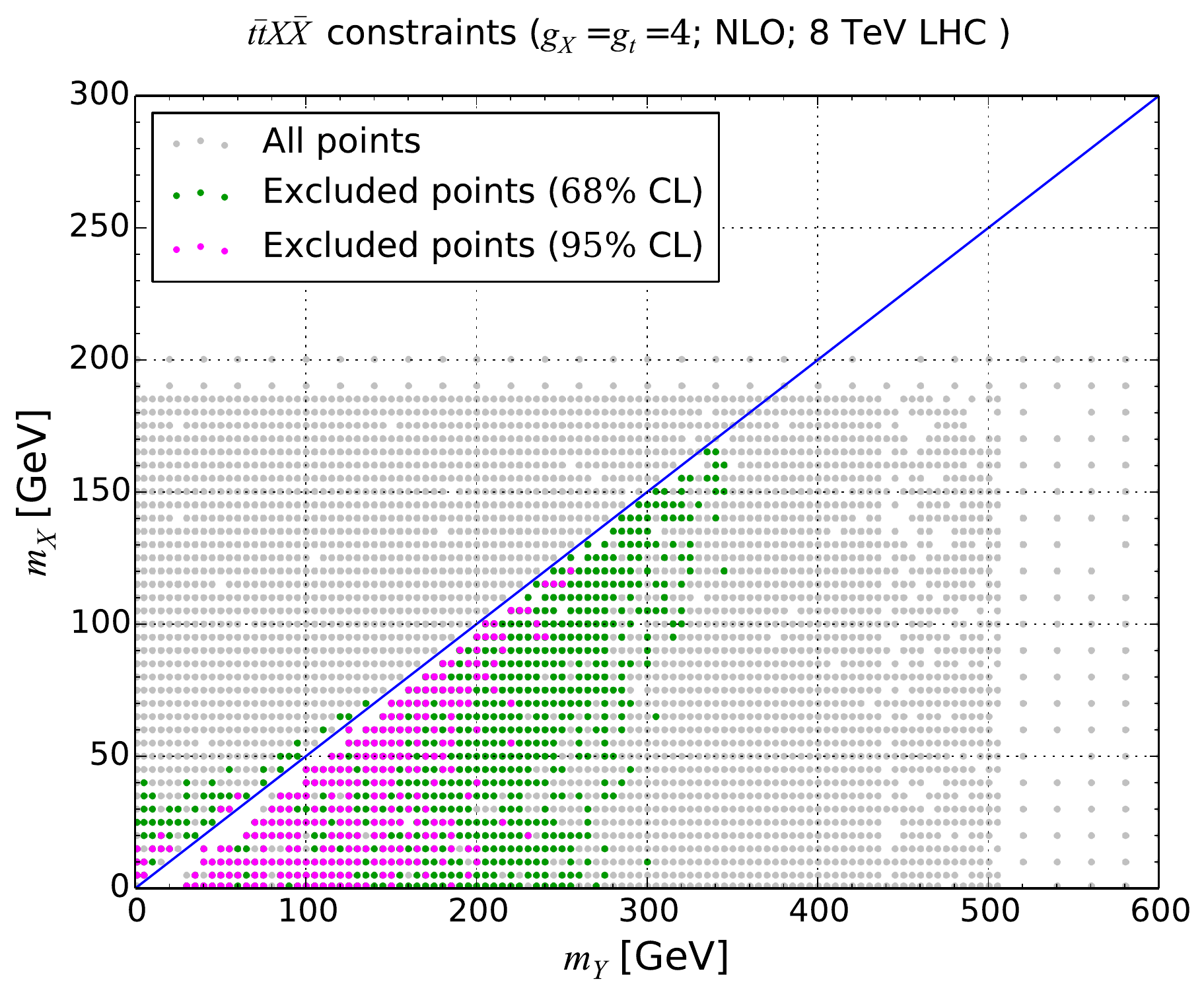}
  \caption{LHC Run~1 sensitivity to the considered model (with $g_t=g_X=4$)
    based on the CMS dark matter analysis in the top-antitop plus missing
    energy channels. Signal simulations are achieved at LO (left) and NLO
    (right) in QCD.}
  \label{fig:ttmet}
\end{figure}

Top-philic dark matter can be explored at colliders through a signature
comprised of a top-quark pair and missing energy. We focus on the corresponding
Run~1 CMS analysis~\cite{Khachatryan:2015nua}. In order to examine the impact of
the QCD corrections on the sensitivity, we first fix
the new physics couplings $g_t=g_X=4$ and scan over the mediator and dark matter
masses $m_Y$ and $m_X$.
The results are shown in Figure~\ref{fig:ttmet} when LO (left panel) and NLO
(right panel) signal simulations are considered, and mass configurations
excluded at the 68\% (green) and 95\% (magenta) confidence level (CL) are
indicated. All excluded points (at the 95\% CL) lie in the triangular low-mass
region where the mediator is heavy enough to resonantly decays into a pair of
dark matter particles, which corresponds to the region with the largest signal
cross section. Scenarios featuring mediator masses ranging up to about
200--250~GeV are excluded, especially if the mediator decays close to threshold
($m_Y\sim2m_X$). Further from threshold, the mediator width can become large by
virtue of the large coupling choice, so that the excluded region is
not exactly triangular and the signal fiducial cross section acquires a dark
matter mass dependence.

\begin{table}
  \begin{center} \renewcommand\arraystretch{1.3}
  \begin{tabular}{clllll}
   & ($m_Y$, $m_X$)  & $\sigma_{\textrm{LO}}$ [pb] & CL$_{\rm LO}$ [\%]
     & $\sigma_{\textrm{NLO}}$ [pb] & CL$_{\rm NLO}$ [\%] \\
   \hline
   I &   (150, 25)~GeV
     & 0.658$^{+34.9\%}_{-24.0\%}$ & 98.7$^{+0.8\%}_{-13.0\%}$
     & 0.773$^{+6.1\%}_{-10.1\%}$ & 95.0$^{+2.7\%}_{-0.4\%}$ \\[1mm]
   II &  (40, 30)~GeV
     &  0.776$^{+34.2\%}_{-24.1\%}$ & 74.7$^{+19.7\%}_{-17.7\%}$
     & 0.926$^{+5.7\%}_{-10.4\%}$ & 84.2$^{+0.4\%}_{-14.4\%}$ \\[1mm]
   III & (240, 100)~GeV
     & 0.187$^{+37.1\%}_{-24.4\%}$ & 91.6$^{+6.4\%}_{-18.1\%}$
     & 0.216$^{+6.7\%}_{-11.4\%}$ & 86.5$^{+8.6\%}_{-5.5\%}$ \\[1mm]
  \end{tabular} \renewcommand\arraystretch{1.0}
  \caption{Benchmark scenarios used to determine the NLO impact on
    the sensitivity of the Run~1 CMS search for dark matter in the top-antitop
    plus missing energy channel. LO and NLO cross sections are
    shown together with the CLs exclusions, all results including scale
    uncertainties.}
  \label{tab:ttmet}
  \end{center}
\end{table}

As NLO corrections have a mild effect on the shape of the relevant signal
distributions~\cite{Backovic:2015soa}, their impact is entirely driven by
changes in the signal rates. We observe that in the low-mass region (where the
$K$-factor is found to be of about 1.10), the exclusion
contours do not feature any significant effect. For larger $m_Y$ values,
the $K$-factors are of about 1 and the exclusion region is identical at LO and
NLO. NLO corrections are however expected to significantly reduce theoretical
uncertainties. In order to study this, we select the three benchmark scenarios
presented in Table~\ref{tab:ttmet} together with the corresponding LO and NLO
signal cross sections and the CLs exclusions. We additionally indicate the scale
uncertainty bands that have been obtained by
varying the central scale, set to the average transverse mass of all final state
particles, by a factor of 2 up and down. Using NLO predictions leads to
a significant reduction of the uncertainties in the total cross section,
compared with the LO case, which propagates down to the CLs exclusions. NLO
predictions therefore allow us to draw more reliable conclusions on the
viability of a parameter space point.

\begin{figure}
  \center
  \includegraphics[width=0.48\textwidth]{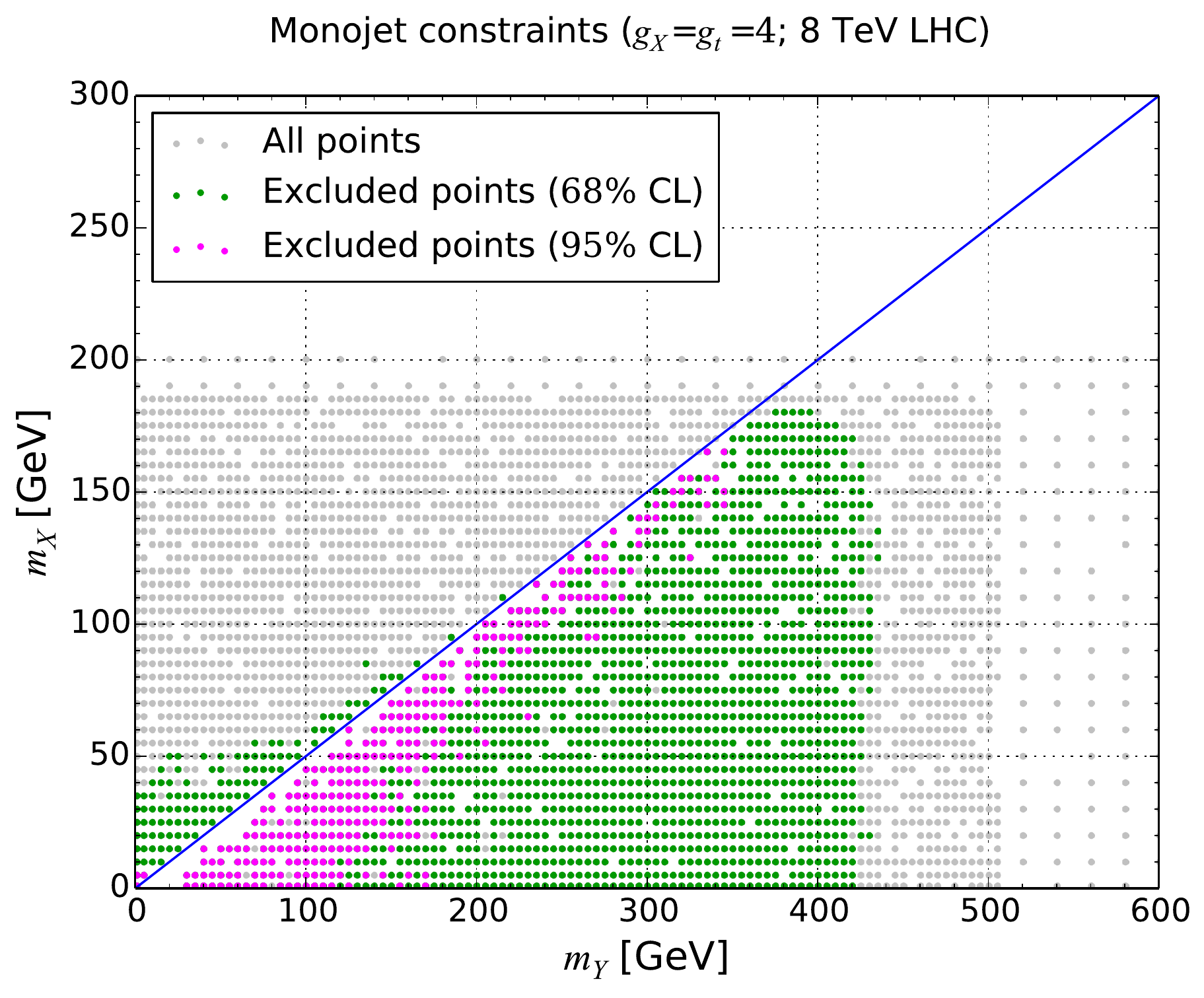}
  \includegraphics[width=0.48\textwidth]{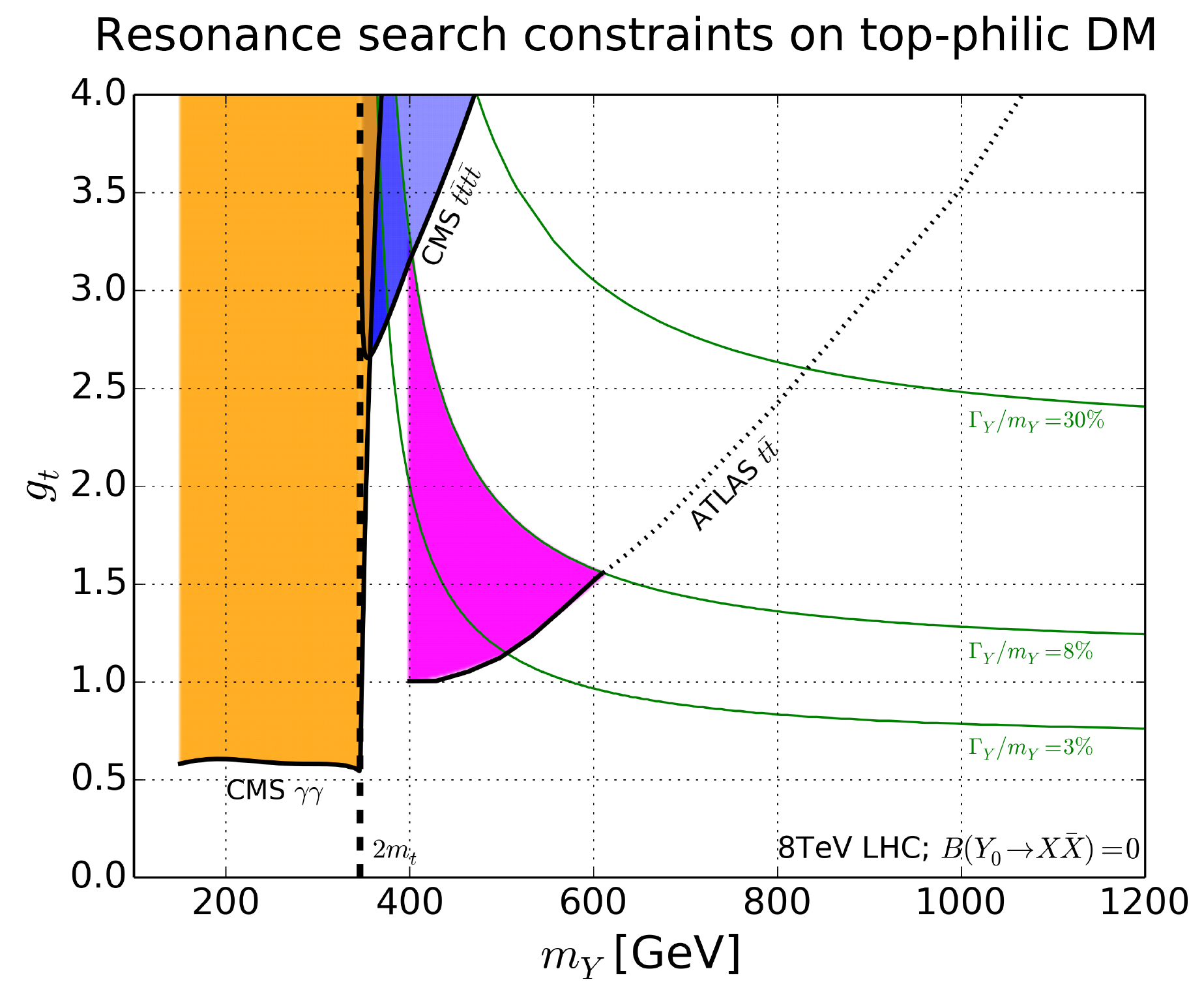}
  \caption{Left: LHC Run~1 monojet constraints on the considered
    top-philic dark matter model as obtained by reinterpreting the Run~1
    CMS monojet search. Right: LHC Run~1 constraints issued from
    the reinterpretation of new physics searches in the diphoton (orange),
    $t\bar t$ (magenta) and $t\bar tt\bar t$ (blue) modes. Information on
    the mediator width to mass ratio is represented by green curves.}
  \label{fig:monojet_nomet}
\end{figure}

Mono-$X$ signals, in which a pair of dark matter particles is produced together
with a Standard Model particle $X$, can also be relevant for obtaining bounds on
dark matter models. Amongst all these searches, we consider the monojet one
given the relative magnitude of the strong and electroweak couplings. As an
example, we use the Run~1 CMS monojet analysis~\cite{Khachatryan:2014rra} and
signal simulations at LO (as loop-induced). In our recasting procedure, we
conservatively select the signal region driving the strongest bound. The
results are shown in the left panel of Figure~\ref{fig:monojet_nomet} for
$g_t=g_X=4$, in which we represent scenarios excluded at the
68\% (green) and 95\% (magenta) CL. Most excluded scenarios lie
again in the triangular low-mass region of the parameter space, in which the
mediator resonantly decays into a pair of dark matter particles. Except for the
small subset of points excluded at the 68\% CL for setups where $m_Y<2 m_X$, the
boundaries of the excluded region are derived from the significant reduction of
the monojet fiducial cross section that rapidly falls with $m_Y$ once the top
threshold is crossed. The LHC Run~1 is therefore not sensitive to mediator
heavier than 500~GeV. In the offshell $m_Y<2 m_X$ regime, jets are
harder~\cite{Mattelaer:2015haa}, so that the signal selection efficiency is
larger and such scenarios can be reached. The monojet search is overall found to
be more constraining than the top-antitop plus missing energy one, especially
for heavier mediator thanks to the larger monojet cross section.

\subsection{Searches without missing transverse energy}
Dark matter models can also be constrained by resonance searches when the new
physics resonance decays into a pair of Standard Model particles, as mediators
can sometimes decay back into the Standard Model sector. Dijet and diphoton
resonance searches could in principle constrain the considered model. However,
due to the double-loop suppression
stemming from the mediator couplings to gluons and photons, new physics
contributions to dijet and diphoton production are only relevant for $m_Y <2\
{\rm min}(m_X, m_t)$, \ie~when the mediator cannot decay into top quarks and
dark matter. Due to the large QCD background and trigger reasons, LHC dijet
resonance searches target high invariant-mass dijet systems, so that the lowest
mediator mass that could be reached by Run~1 data is of about 500~GeV, the
corresponding cross section bound being 10~pb~\cite{Khachatryan:2016ecr}.
This consequently yields no constraint on the model. In contrast,
despite the small mediator branching ratio into a diphoton final
state, diphoton probes can potentially constrain the model due to the low level
of associated backgrounds. In the right panel of Figure~\ref{fig:monojet_nomet},
we present, in the $(m_Y, g_t)$ plane, the constraints derived from the results
of the Run~1 CMS diphoton search~\cite{Khachatryan:2015qba} that investigates
resonance masses ranging from 150 to 850~GeV. The bounds are obtained by
comparing the experimental limits with predictions for single mediator
production at the next-to-next-to-leading-order (NNLO)~\cite{%
Heinemeyer:2013tqa}.

For scenarios with mediator masses above the top threshold, one relies
instead on signatures featuring a $t\bar{t}$ resonance. In our setup,
loop-induced single mediator production can enhance top-antitop production, in
particular when there is a large coupling hierarchy ($g_t\gg g_X$) or mass
hierarchy ($2m_t<m_Y<2m_X$). We extract constraints from the results of the
Run~1 ATLAS $t\bar t$ resonance search~\cite{Aad:2015fna} that restricts the new
physics contributions to the $t\bar t$ cross section as a function of the
mediator mass. Comparing with NNLO signal cross section predictions,
we show the obtained constraints in the right panel
of Figure~\ref{fig:monojet_nomet} (magenta) in the $(m_Y, g_t)$ plane, dark
matter being again decoupled. Mediators with masses lying in the [400,600]~GeV
range are found to be reachable for $g_t \in [1, 4]$. The exact form of the
excluded region depends on $m_Y$ and on the validity of the narrow-width
approximation that must be enforced to recast the ATLAS limits (green curves).
In addition, mediators can also be produced in association with a top-antitop
pair, which yields a four-top signal after decay. Large new physics four-top
contributions are forbidden~\cite{Beck:2015cga}, and recasting the limits
extracted from the CMS analysis of Ref.~\cite{Khachatryan:2014sca} leads to the
weak blue excluded region in Figure~\ref{fig:monojet_nomet}, as resulting from
the steeply decreasing (NLO) signal cross section with $m_Y$.

\section{Summary - the collider-cosmology complementarity}\label{sec:combine}
\begin{figure}
  \centering
  \includegraphics[width=.48\columnwidth]{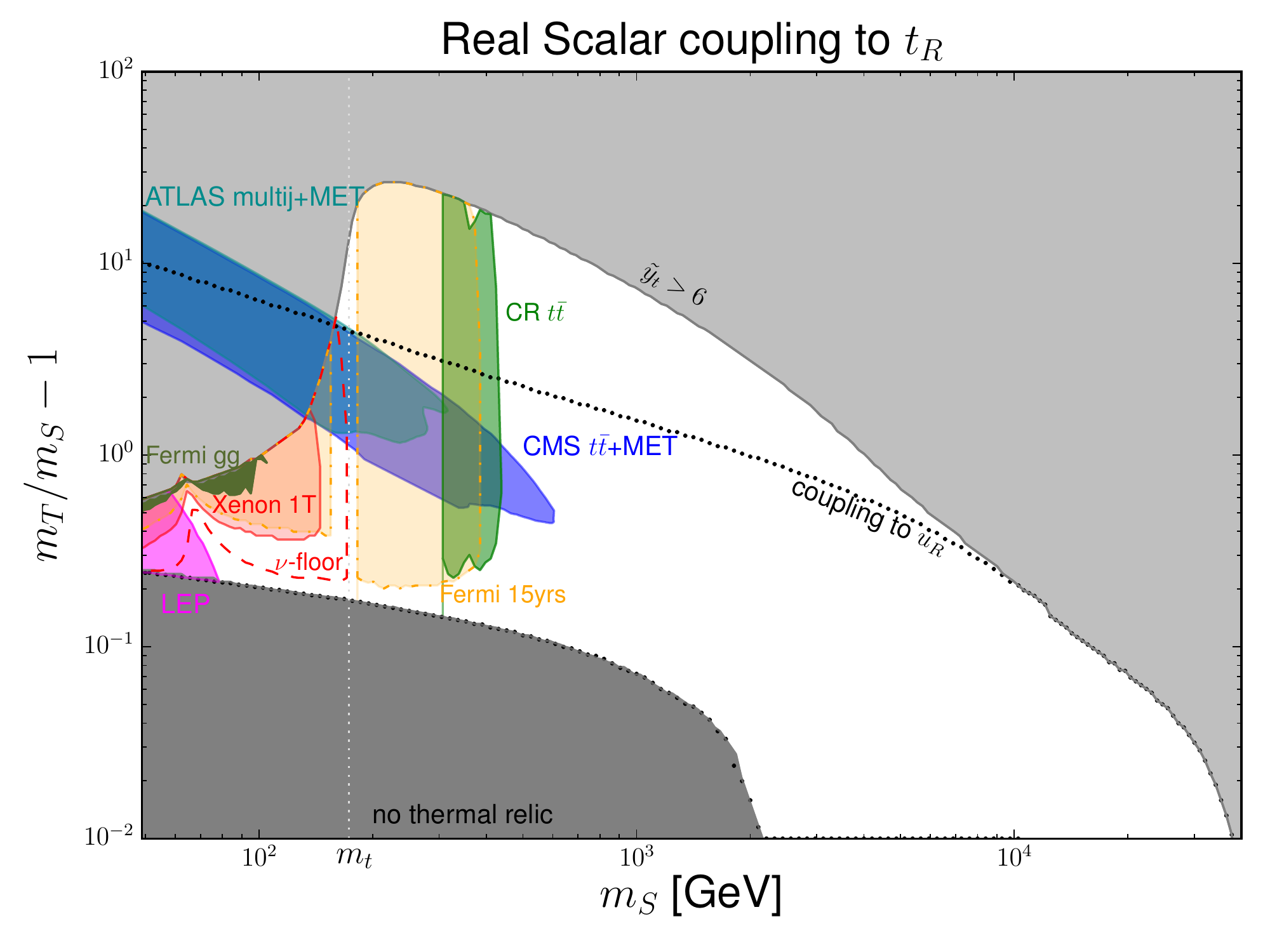}
  \caption{Phenomenologically-viable region of the top-philic dark matter model
    of Section~\ref{sec:cosmo}, presented in the $(m_S, r-1)$ plane. The viable
    region of the parameter space (non-grey regions) is obtained by fixing the
    $\tilde y_t$ coupling to reproduce the observed relic density. We include
    constraints from direct and indirect dark matter detection as well
    as from colliders.}
  \label{fig:all}
\end{figure}

We now go back to the model of Section~\ref{sec:cosmo} and jointly apply all
the constraints discussed in this paper. Our findings, reported in
Figure~\ref{fig:all}, reflect the nice complementarity of cosmological and
collider (LEP as well as LHC Run~1 and 2) constraints on the considered dark
matter model. We compare the impact of experimental searches from varied origins
by confronting their results to predictions including NLO QCD corrections, the
latter being crucial for a good theoretical control.
The red region is excluded by dark matter direct detection searches from the
Xenon 1T~\cite{Aprile:2017iyp} experiment, and the region inside the red dashed
line is in principle testable by future direct detection searches as above
the neutrino floor. The dark and light green regions are excluded by Fermi-LAT
gamma-ray and cosmic-ray constraints on low and large dark matter masses
respectively~\cite{Ackermann:2015zua,Cuoco:2017iax}, while the orange area will
be testable after 15 years of Fermi-LAT running~\cite{%
Charles:2016pgz}. Finally, collider searches in the top-antitop plus missing
energy channel at LEP~\cite{Abbiendi:2002mp} and at the LHC~\cite{%
Sirunyan:2017leh,CMS:2017odo} are presented as the magenta and light blue
contours
respectively, and mono-$X$ bounds~\cite{Aaboud:2016tnv,Aaboud:2016zdn} are given
in dark blue.

Our analysis reveals that despite the complementarity
between the different sets of constraints, only a small fraction of the
parameter space is currently tested. The prospects of all cosmological probes
however demonstrate that the most fruitful strategy to further test the model on
the longer run would be to increase the energy reach at colliders.

\section*{Acknowledgments}
I am thankful to the organisers of the Kruger 2018 conference for the invitation
and for putting such an inspiring meeting in place. I also warmly thank Alan
Cornell for his hospitality at U.~Witwatersrand. This study has been partly
supported by French state funds managed by the Agence Nationale de la Recherche
(ANR) in the context of the LABEX ILP (ANR-11-IDEX-0004-02, ANR-10-LABX-63).

\section*{References}

\bibliographystyle{iopart-num}
\bibliography{biblio}

\end{document}